\documentclass[a4paper,preprintnumbers,floatfix,superscriptaddress,prx,twocolumn,showpacs,notitlepage,longbibliography]{revtex4-1}
\usepackage[utf8]{inputenc}
\usepackage[T1]{fontenc}
\usepackage[sc,osf]{mathpazo}
\usepackage{amsmath, amsthm, amssymb,amsfonts,mathbbol,amstext}
\usepackage{graphicx}
\usepackage{dcolumn}
\usepackage{bm}
\usepackage{bbm}
\usepackage{hyperref}
\usepackage{mathtools}
\usepackage{comment}
\usepackage{color}
\usepackage{multirow}
\usepackage{diagbox}
\usepackage{float}
\usepackage[normalem]{ulem}
\usepackage{verbatim}

\def\1{\mathbf{1}}
\def\0{\mathbf{0}}

\newcommand{\mean}[1]{\left\langle #1 \right\rangle}

\newcommand{\ie}{{\it{i.e.~}}}

\newcommand{\processnext}[1]{%
  \ifx\listfinish#1\empty\else\listact{#1}\expandafter\processnext\fi}

\DeclareGraphicsExtensions{.pdf,.png,.jpg}

\begin{document}

\title{Satellite-based photonic quantum networks are small-world}

\author{Samura\'i Brito}
\affiliation{International Institute of Physics, Federal University of Rio Grande do Norte, 59070-405 Natal, Brazil}
\author{Askery Canabarro}
\affiliation{International Institute of Physics, Federal University of Rio Grande do Norte, 59070-405 Natal, Brazil}
\affiliation{Grupo de F\'isica da Mat\'eria Condensada, N\'ucleo de Ci\^encias Exatas - NCEx, Campus Arapiraca, Universidade Federal de Alagoas, 57309-005 Arapiraca-AL, Brazil
}
\author{Daniel Cavalcanti}
\affiliation{ICFO-Institut de Ciencies Fotoniques, The Barcelona Institute of
Science and Technology, 08860 Castelldefels (Barcelona), Spain}
\author{Rafael Chaves}
\affiliation{International Institute of Physics, Federal University of Rio Grande do Norte, 59070-405 Natal, Brazil}
\affiliation{School of Science and Technology, Federal University of Rio Grande do Norte, 59078-970 Natal, Brazil}

\begin{abstract}
Recent milestone experiments establishing satellite-to-ground quantum communication are paving the way for the development of the quantum internet, a network interconnected by quantum channels. Here we employ network theory to study the properties of the photonic networks that can be generated by satellite-based quantum communication and compare it with the optical-fiber counterpart. We predict that satellites can generate small-world networks, implying that physically distant nodes are actually near from a network perspective. We also analyse the connectivity properties of the network and show, in particular, that they are robust against random failures. This puts satellite-based quantum communication as the most promising technology to distribute entanglement across large distances in quantum networks of growing size and complexity.
\end{abstract}

\maketitle

\section{Introduction}
Quantum networks, distant nodes interconnected by quantum channels, are ubiquitous in quantum information science. Not only they can enhance our communication capabilities \cite{densecoding} but also allow for a fundamentally secure quantum cryptography \cite{Gisin_2002}, the execution of quantum teleportation \cite{teleportation}, the establishment of nonlocal correlations \cite{Brunner_2014} and other applications such as clock synchronisation \cite{ClockSinc} and quantum computation on the cloud \cite{Broadbent_2009,PhysRevA.96.012303}.  
The two most promising infrastructures to build quantum networks are optical fibers and satellites \cite{simon2017towards}. Recent experimental advances \cite{valivarthi2016quantum,wengerowsky2019entanglement,Bedington2017,Yin1140,PhysRevLett.120.030501} allowed for quantum communication and the sharing of quantum entanglement through large distances, paving the way for undergoing development of the quantum internet \cite{kimble2008quantum, wehner2018quantum,HuangLiang}. 

Within this context, it becomes crucial to understand the network properties of a quantum internet generated by these two technologies. For instance, the connectivity of the network (\ie if all nodes belong to the same network or if there exist isolated islands of nodes) tells us  if it is possible to transmit information across the whole network. The network distance between nodes tell us how many entanglement swaps are needed if one wants to distribute entanglement between these nodes. Finally, the network robustness (\ie, how many nodes must be removed from the network until it breaks apart) informs us about how resilient the network is under local failures. In a recent study \cite{samurai1} we have shown that an optical-fiber-based network requires a very small density of nodes in order to produce fully connected photonic networks. However, the typical distances between nodes increase in a power-law relation with the number of nodes, \ie the size of the network, meaning that it does not lead to the so-desired small-world property \cite{small1,small2,small3,small4,small5}. 

Here we employ network theory \cite{barabasi2016network} to study the properties a satellite-based photonic network. Similarly to an optical-fiber-based quantum internet (OFBQI), the satellite-based quantum internet (SBQI) displays a transition from a disconnected to a connected network with respect to the density of nodes in the network. However, differently from an OFBQI, we observe the existence of hubs in a SBQI, \ie nodes with a high number of connections, a phenomenon typically associated with scale-free networks \cite{barabasi2016network}. Consequently, the SBQI leads to the small-world property, implying that very few entanglement swappings are necessary to distribute entanglement between any two nodes in the network. Moreover, similarly to the actual internet \cite{albert2000error,PhysRevLett.85.4626}, the SBQI also displays a significant robustness against random failures in the network. On the down side, given the existence of hubs, it is less robust against targeted attacks, thus showing that highly connected nodes have to be specially protected in order to maintain the quantum network operative.

\begin{figure*}[t!]
\begin{center}
\includegraphics*[width=\textwidth]{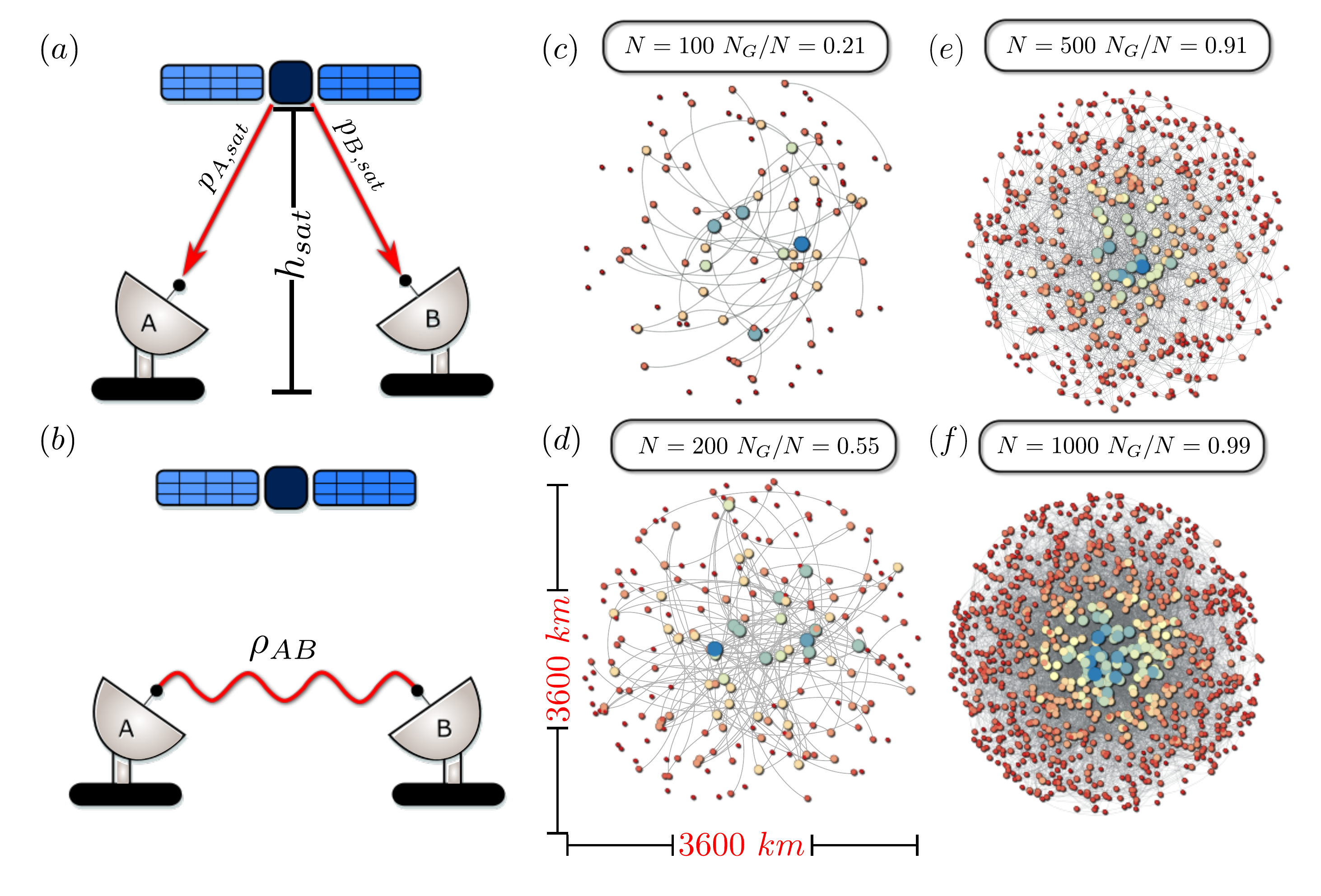}
\end{center}
\caption{\textbf{Satellite-to-ground quantum communication and samples of the quantum internet}. $(a)$ A satellite produces an entangled pair of photons and sends one photon to each station. The probability that each photon arrives at the destination is given by $p_{i, sat}$, and $(b)$ a link between the two ground stations are formed when both photons arrive. (c) to (f) display samples from the quantum internet. The gray edges represent the quantum link created by the satellite between two distant parts. The bigger (smaller) and bluish (reddish) the nodes, more (less) connected they are. By increasing the number of nodes $(c)$ $N=100$, $(d)$ $N=200$, $(e)$ $N=500$ and $(f)$ $N=1000$ in a fixed area of radius $R = 1800$ km, using $n_p = 50$ photons, the giant cluster will appear,  \ie the number of nodes within the largest cluster $N_{G}$ is of the order of the total number of nodes $N$.}
\label{sat_model}
\end{figure*}

\section{Network model for the satellite based quantum internet}

Formally, a network model is defined by a set of $N$ nodes being connected by edges according to a given probabilistic rule. In our model (see Fig. \ref{sat_model}) we assume that there is a satellite covering a disk of radius $R$, capable of connecting any two nodes within this area.
More precisely, an entangled pair of photons are generated at the satellite and sent to each pair of nodes, via a down-link channel. The probability that a photon sent from the satellite arrives at site $i$ has been analyzed in \cite{Bonato_2009} and can be described by
    \begin{eqnarray}
        p_{i,sat} &=& \eta_0(1 - e^{2R^2_{rec}/(w^i_{LT})^2}),
    \end{eqnarray}
where $ w^i_{LT} \simeq 0,25\times 10^{-5}d_{i,sat}$ (for a downlink) is the long-term beam width and depends on the distance $d_{i,sat}$ between the site and the satellite; $\eta_0 \approx 0.1$ is an empirical factor~\cite{teleportation2} comprising the detection efficiency, the pointing losses and the atmospheric attenuation; $R_{rec}$ is the radius of the telescope receptor (in our simulations we consider  $R_{rec}=0,75m$). The distance $d_{i,sat}$ between site $i$ and the satellite will depend on the position of the site within the coverage radius and the altitude $h_{sat}$ of the satellite, that for the Micius satellite \cite{gibney2016one}, used as reference here, is $h_{sat} = 500$ km. Furthermore, we assume that the position of the satellite is fixed and positioned in the center of the disk. Finally, we will consider that the satellite can send $n_p$ entangled pairs as an attempt to generate a link. A successful link is then established between sites $i$ and $j$ if at least one out of $n_p$ entangled pairs generated at the satellite arrive in the stations. This connection will happen with probability
    \begin{eqnarray}
    \Pi_{i j} &=& 1- (1 - p_i p_j)^{n_p},
\end{eqnarray}
with $p_{i} \equiv p_{i,sat}$ (similarly for $p_j$).

\begin{figure*}[t!]
\begin{center}
\includegraphics[width=\textwidth]{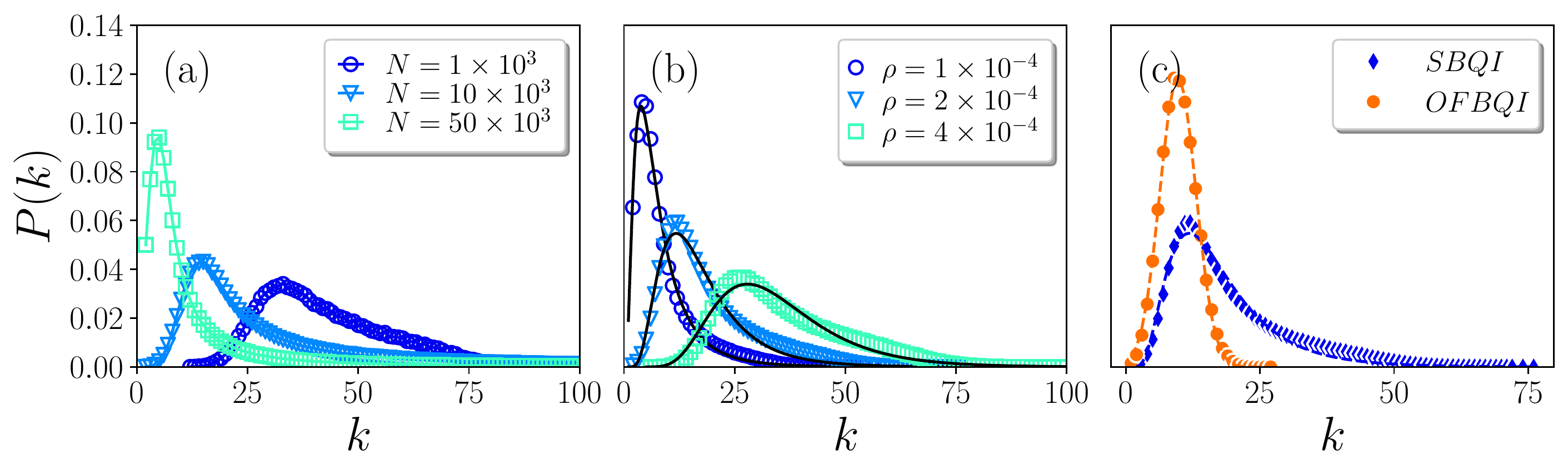}
\end{center}
\caption{\textbf{Connectivity distribution of SBQI model and a comparison with the OFBQI.} Connectivity distribution of a SBQI model for $(a)$ different values of $N$, setting $\rho = 5\times10^{-3}$ and $(b)$ several values of $\rho$, setting $N=1000$. In solid lines, the log-normal distribution \eqref{lognormal}  parameterized by $\mu$ and $\sigma$, fitting almost perfectly the distribution generated by the model. $(c)$ Comparison between the satellite (red diamond) and the optical fibers (blue circle) networks for $N=1000$ and $\rho=2\times 10^{-4}$ $(R \simeq 1261\;km)$. The appearance of hubs in the satellite model becomes evident, according to the fat-tailed behaviour. For the satellite model $n_p = 50$ and for the fiber one $n_p = 1000$, that is, we are allowing 20 times more photons for the optical-fiber model to establish a quantum link, an important advantage of the satellite implementation.}
\label{connectivity}
\end{figure*}

Within this framework, our model for the SBQI can be constructed in three steps:

\begin{enumerate}
    \item $N$ nodes are uniformly distributed in a disk of radius $R$ (km); 
    \item Compute the Euclidean distance  $(d_{i, sat}, d_{j,sat})$ between $i$ and $j$ and the satellite, for all pair of nodes $i$ and $j$;
    \item For each pair $(i,j)$ randomly sample a number uniformly distributed in the range $0\leq r\leq 1$. If $r \leq \Pi_{i j}$ the sites are connected, otherwise they are not.
\end{enumerate}

%\sam{An important point here is the uniform distribution of the nodes on the disk. If we would replace our nodes placement (uniformly at random) by a more regular network (like a 2D grid) our results would not change. That follows from the fact that the density of nodes in both models are uniform. However, if one consider a non-uniform distribution of nodes one can expect qualitative and quantitative differences. This non-uniformity could arise, for instance, from the fact that big cities will concentrate more nodes and less densely populated or rural areas will have few nodes. We expect that this situation will lead to the formation of what is called communities in network science. That is, the nodes that are closer geographically will have a higher chance of being connected.} 

To calculate the relevant properties of our network model in a statistically relevant manner, we used the standard Monte Carlo method until $1000$ steps to generate different instances of the SBQI model. In our simulations we have employed different values of the parameter $n_p$, observing that it does not change the qualitative properties of our model ($n_p=50$ was chosen as a reference for the rest of the paper, unless stated otherwise).  Examples of the generated networks are shown in Fig.~\ref{sat_model} (c)-(f). As it can be seen, even a few nodes in a large area can generate a network in a connected phase,  \ie $N_{G}\sim N$, where the variable $N_G$ is defined as the number of nodes in the largest connected component (the giant cluster), and $N$ is the size of the entire network, see Fig. \ref{sat_model} (e) and (f). Another interesting feature is that, as the number of nodes increases, the most connected sites will naturally appear in the center of the disk (directly under the satellite). The further away from the center, the less connected are the nodes, also meaning that the hubs are likely to appear in the center of the network area.

\section{Satellite-based quantum communication generates small-world networks}

A crucial property of a network that dictates most of its qualitative and quantitative properties is the connectivity distribution $P(k)$, \ie the probability of finding a node with $k$ connections. As shown in Fig. \ref{connectivity}, we find that $P(k)$ can be well fitted by a log-normal distribution
\begin{equation}
\label{lognormal}
P(k) = \frac{1}{k\sigma \sqrt{2\ \pi}} \exp\left[\frac{-(\ln(k) - \mu^2)^2}{2\sigma^2}\right],    
\end{equation}
that depends on the parameters $\mu \equiv \ln\left[\langle k \rangle^2/\sqrt{\langle k^2 \rangle}\right]$ and $\sigma \equiv \sqrt{\ln\left[ \langle k^2 \rangle / \langle k \rangle^2 \right]}$, where $\mean{k}=\sum_i k_i/N$ and $k_i$ is the number of connections that node $i$ has.

A comparison between the connectivity distribution of the SBQI and OFBQI highlights a few important differences. The OFBQI is governed by a Poissonian distribution, meaning that most of the nodes will have a connectivity close to $\mean{k}$ with deviations that become exponentially smaller as the network size increases \cite{samurai1}. On the contrary, the SBQI being governed by a log-normal distribution implies that there will be a non-negligible probability of finding nodes with a considerably high number of connections, which accounts for the fat-tail behavior seeing in Fig. \ref{connectivity}. That is, the SBQI leads to the existence of hubs, a characteristic trait of scale-free networks describing many networks \cite{Albert1999,10.2307/26060284,10.1145/316194.316229,Redner1998,Jeong2000,Yook13382}. For a long time, real-world networks were often claimed to be scale-free and thus governed by a power law distribution. Recent works have shown robust evidences, however, that strongly scale-free structure is empirically rare:  for most networks the log-normal distributions fit the data as well or better than power-laws \cite{Broido2019}.

\begin{figure*}[t!]
\begin{center}
\includegraphics[width=0.75\textwidth]{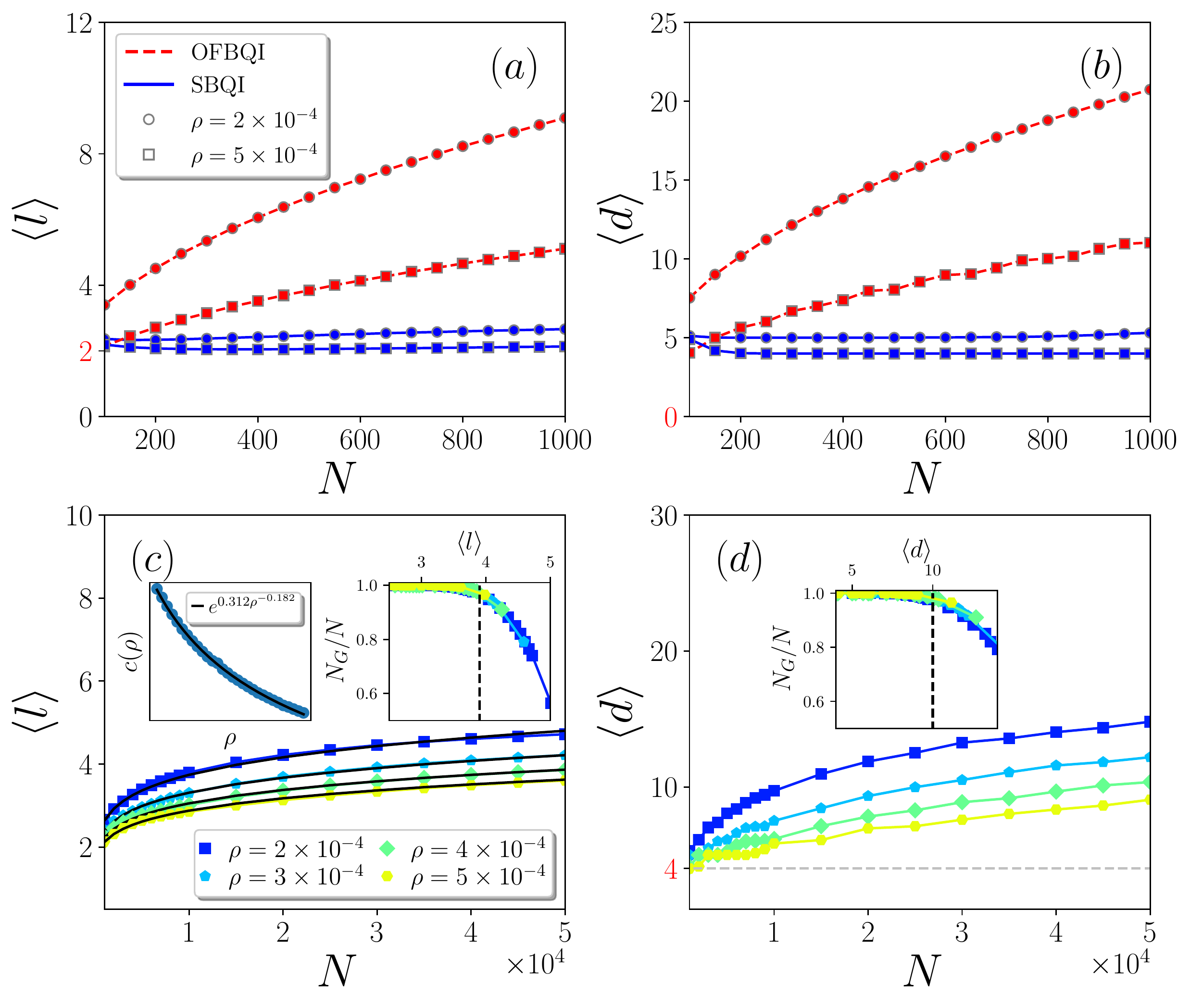}
\end{center}
\caption{\textbf{Comparison between SBQI and OFBQI networks.} Average shortest path $\langle l \rangle$ and average diameter $\langle d \rangle$ as function of $N$ for different values of $\rho \equiv N/\pi R^2$. Here we are considering only the giant cluster, once that by definition two disconnected nodes has $\langle l \rangle = \infty$.  $(a)$ Comparison of $\langle l \rangle$ between OFBQI (red dashed line) and SBQI (blue straight line) for $\rho = 2.0 (\times 10^{-4})$ (circles) and $5.0(\times 10^{-4})$ (squares). $(b)$ Comparison of the diameter of the network $\langle d \rangle$ for OFBQI and SBQI models for the same parameters as before, once more highlights the advantage of the satellite model. Plots (c) and (d) show that $\langle l \rangle$ and $\langle d \rangle$ remain small for the SBQI, even for very large $N$ (also considering different density values $\rho$). The inset plots in (c) (right) and (d) show the relative size of the giant cluster as a function of the average shortest path and diameter respectively. In the connected regime of the network ($N_G \sim N$) the shortest path is at most $\langle l \rangle_{max} \sim 4$ and $\langle d\rangle_{max} \sim 10$, independently of $N$ and $\rho$. The black straight lines in Fig.~\ref{fig:d_mc}(c) are given by the equation $c(\rho)\ln(N)/\ln(\langle k\rangle/\rho)$ with $c(\rho)=e^{0.312\rho^{-0.182}}$, the dependence of $c(\rho)$ as a function of $\rho$ is displayed in the left inset plot in (c).}
\label{fig:d_mc}
\end{figure*}

\begin{figure*}[t!]
\begin{center}
\includegraphics[width=0.75\textwidth]{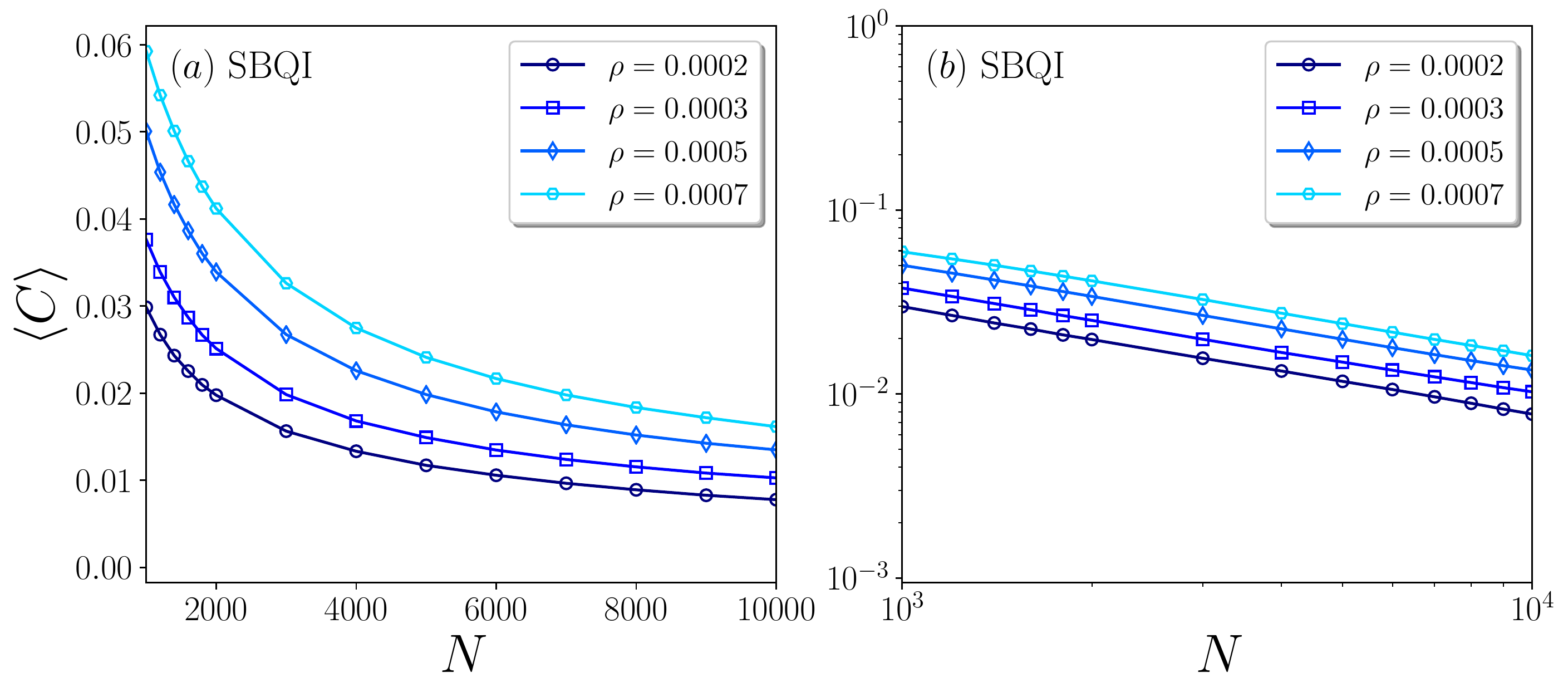}
\end{center}
\caption{\textbf{Average clustering coefficient.} $(a)$ $\langle C \rangle$ for SBQI model as a function of $N$ for several values of $\rho$. $(b)$ As can be seen, $\langle C \rangle$ nearly decreases as a power of $N$. The simulations were done for $1000$ realizations for each value of $N$ with $N_{max} = 10000$. To be sure if $\langle C \rangle\sim N^{-\alpha}$ would be necessary to run for higher values of $N$.}
\label{fig:cmed}
\end{figure*}

We also notice that it is intuitive to expect that the current model leads to a connectivity distribution behavior that is in between the random networks (given by a Poissonian) and scale-free networks (given by a power-law). This is because, while in the first every pair of nodes have the same probability of receiving a connection, in the later nodes that have more links are more likely to receive more connections. However, in the current model, although there is no preferential attachment, the distance between the nodes and the satellite changes the probability to receive links: the nodes lying below the satellite will naturally display more connections than the ones lying in the extremes of the satellite coverage area. Thus, although we expect the appearance of hubs, they will be highly concentrated in the central area. 

A consequence of the existence of hubs is to decrease the typical network distance between nodes. This can be checked by looking at how the average shortest path $\mean{l}$ of the network (the average being performed over all pairs of nodes as well as over all samples of the networks) scales with the network size. To compute the average shortest path length $\langle l \rangle$ we use the standard definition in the network science literature, given by
\begin{equation*}
    \langle l \rangle  = \displaystyle \frac{2}{N(N-1)} \sum_{i<j}d_{ij},
\end{equation*}
where $d_{ij}$ is the shortest path between the sites $i$ and $j$. Thus $\langle l \rangle$ is the average of the shortest paths between all pairs of nodes in the network. As standard in the literature, we only compute $\langle l \rangle$ for the sites belonging to the giant cluster, once that $d_{ij} = \infty$ if $i$ and $j$ do not belong to the same subgraph. In a quantum network in the connected phase, any node can become entangled to any other node, via a entanglement swapping over the intermediate nodes. However, due to unavoidable noise, each of those intermediary processes damages the final amount of entanglement between the end nodes, highlighting the relevance of having a small $\mean{l}$. 
Efficient communication networks display the small-world property. Unfortunately, as shown in \cite{samurai1}, the fiber based quantum internet displays no small-world property with $\mean{l} \sim \sqrt{N}$. Here, on the contrary, as shown in Fig. \ref{fig:d_mc}, $\mean{l}$ for the SBQI is governed by the functional $c(\rho)\ln N/\ln[\langle k\rangle/\rho]$ with $c(\rho)=e^{0.312\rho^{-0.182}}$. Thus, the satellite-based quantum networks displays the small-world phenomenon. This property comes from the presence of hubs in the center of the network area that tend to shorten the path between any two nodes in the network. We also analyse the average clustering coefficient defined by $\langle C \rangle = \frac{1}{N}\sum_{i} c_i$, where $c_i$ is the clustering coefficient of the site $i$ given by $c_i=\frac{2n_i}{k_i(k_i - 1)}$ with $n_i$ being the number of edges between the $k_i$ neighbours of the site $i$ and $k_i(k_i - 1)/2$ being total possible number of edges between them. This property is a measure of the local link density, that is, how connected are the neighbours of a given site. Contrary to the traditional small-world model, proposed by Watts and Strogatz \cite{Watts1998}, and differently of OFBQI network,  $\langle C \rangle^{SBQI}$ decreases nearly as power of $N$ (see Fig.~\ref{fig:cmed}). The small-world property ($\langle l \rangle \sim \log N$) with a decreasing clustering coefficient is also observed in the paradigmatic Erdos-Renyi \cite{erdos1959random} or Albert-Barabási \cite{Albert1999} networks.

The satellite network also improves over the optical fiber one when considering the diameter of the network (the greatest distance between any pair of nodes) $\mean{d}$. This quantity gives the maximum number of entanglement swaps needed in order to directly entangle any two nodes. As it can be seen in Fig. \ref{fig:d_mc}(b), using a satellite as a quantum channel generates a final network with a considerably shorter diameter as compared to optical fibers. For instance, choosing $N=1000$ and the coverage area as $R \approx 1260$ km and even by setting $n_p^{OFBQI}=1000$ while $n^{SBQI}_p = 50$, the diameter of the OFBQI network would be greater than the SBQI case ($d_{OFBQI} \sim 20$, $d_{SBQI}\sim 4$).

In summary, using satellites and a few number of nodes one can achieve a fully connected network covering significantly large areas and requiring few entanglement swaps to interconnect any two nodes, a clear advantage in the practical implementation of the quantum internet.

\begin{figure*}[t!]
\begin{center}
\includegraphics[width=\textwidth]{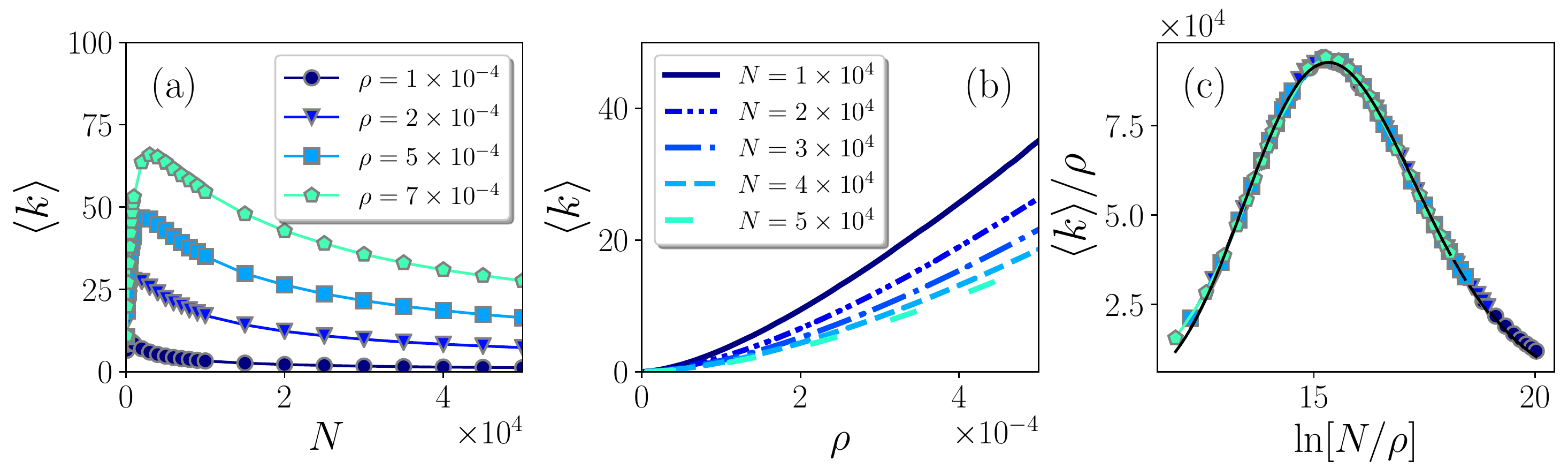}
\end{center}
\caption{\textbf{Average degree of the network} $(a)$, $(b)$ show $\langle k \rangle$ as a function of $N$ and $\rho$, respectively. $(c)$ Eq.~\ref{eq:kmed} (solid black line) fits very well the average degree where by replacing $\langle k \rangle \to \langle k \rangle/\rho$ and $N \to \ln[N/\rho]$, all curves colapse obtaining an universal behaviour independently of $N$ and $\rho$.} 
\label{kavg}
\end{figure*}

\section{Connectivity of the quantum internet}

Another relevant difference between the two models is that, in the fiber model the average connectivity depends linearly on the density of nodes $\rho=N/\pi R^2$ such that $\mean{k}=\alpha \rho$ \cite{samurai1}. We observe that for the SBQI $\langle k \rangle$ has a much more intricate functional dependence, well described by
\begin{eqnarray}
\langle k \rangle = \frac{A(\rho)}{\ln(\pi R^2)\sigma \sqrt{2\pi}}\exp{\left[-\frac{(\ln\ln(\pi R^2) - \mu)^2}{2\sigma^2}
\right]},\label{eq:kmed}
\end{eqnarray}
with $A(\rho) \simeq 4.5\times10^5 \rho + 0.97$, $\mu\simeq2.73$ and $\sigma \simeq 0.126$. To find the expression for $\langle k \rangle$, we generated the model varying the parameters $N$ and $\rho$ and analyzed how $\langle k \rangle$ changed with $N$ fixing $\rho$ and the other way around (see Fig.~\ref{kavg}(a,b)). By rescaling the axis of the Fig.~\ref{kavg}(a), we could find a universal behavior (data collapse) of $\langle k \rangle$. We fit the final data (Fig.~\ref{kavg}(c)) and derived an expression  for $\langle k \rangle$ that describes very well the simulation data. Interestingly, as can be seen in Fig. \ref{kavg} $(a)$, $\langle k \rangle$ has a maximum value (peak) from which the curve decreases. Since $\rho$ is fixed, by increasing $N$ we also increase $R$. This means that when we increase $N$ we adding nodes which are far from the satellite, and will consequently receive less connections, leading to a decrease of $\langle k \rangle$.

\begin{figure*}[t!]
\begin{center}
\includegraphics[width=\textwidth]{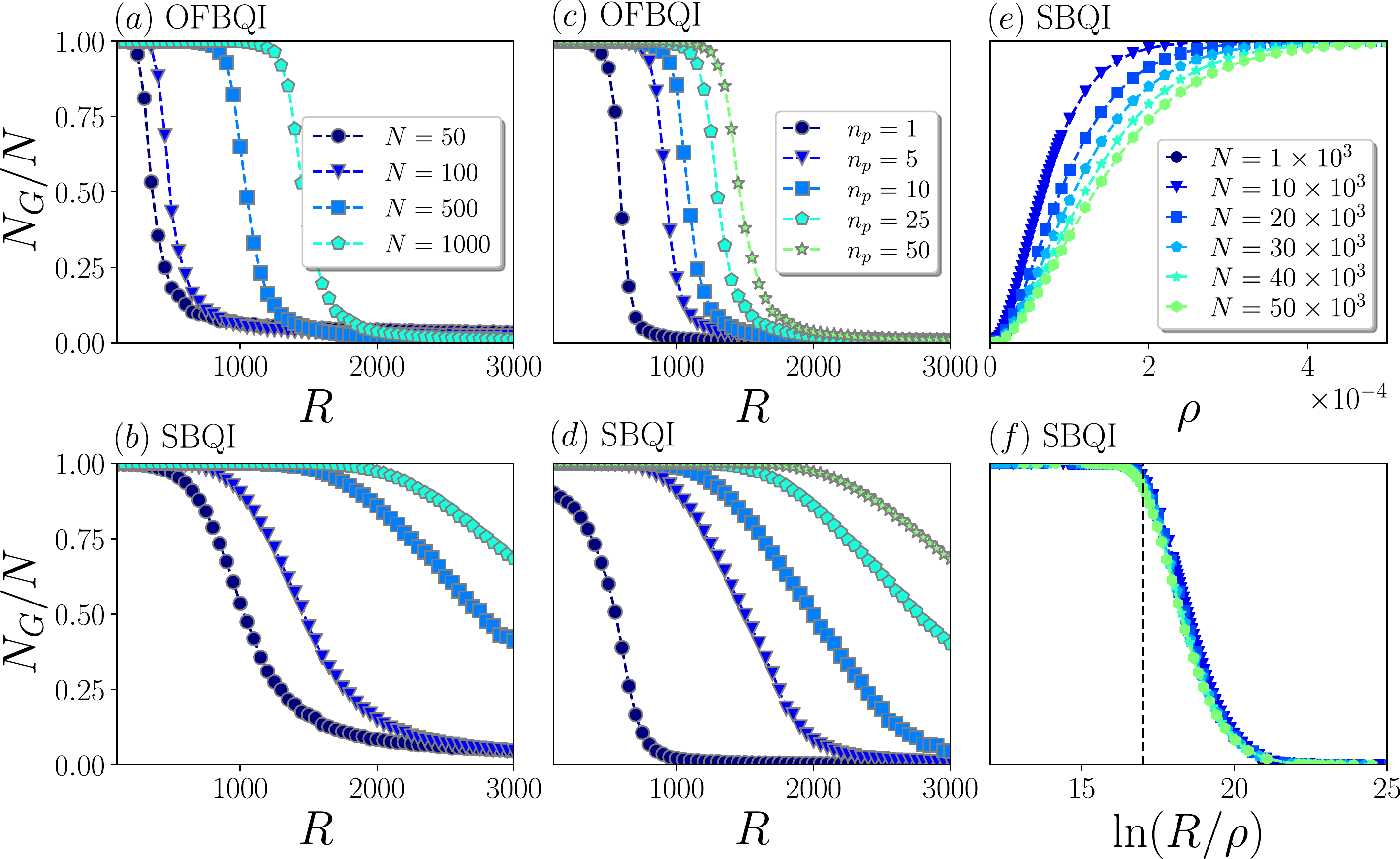}
\end{center}
\label{network}
\caption{\textbf{Relative size of the giant cluster $(N_G/N)$}. Comparison between $(a)$ OFBQI and $(b)$ SBQI models for $n_p = 50$ and $N=50$, $100$, $500$, $1000$. Comparison between the $(c)$ OFBQI  and $(d)$ SBQI models for $N=1000$ and $n_p = 5,10,25,50$. Nicely, the satellite network can be connected in considerably large areas, even with relatively small values of $n_p$ and $N$. In figures $(e)$ and $(f)$ we show the transition between the disconnected and connected phases in the satellite model (for various values of $N$) as a function of $\rho$ and $\ln(R/\rho)$, respectively. As we can see in $(f)$ the network remains connected $(N_G \sim N)$ if $\ln(R/\rho)\lesssim 17$ that is equivalent to $\rho \gtrsim \left[e^{-17}\sqrt{N/\pi}\right]^{2/3}$, indicating that for each $N$ exist a density from which the network becomes connected.}
\label{NG_sat_op}
\end{figure*}

For a communication network to be useful it should have most of its nodes belonging to the same cluster, and not isolated in a few small-size clusters. Models like the paradigmatic Erdos-Rényi random network model \cite{erdos1959random} or the OFBQI \cite{samurai1}, the emergence of the giant cluster is regulated by the average connectivity, such that it happens if $\mean{k}$ is above a critical value. As shown in Fig. \ref{NG_sat_op}, the SBQI also displays a transition from small disconnected clusters to a connected largest cluster.

Notice that the SBQI model generates a connected network in large areas and still keeps the average path small. For instance, for $N=1000$ and $n_p=50$, the SBQI model generates a fully connected network $(N_G \sim N)$, covering an area of radius $R\approx 1800$ km. With the same parameters the OFBQI network would cover a maximum area of radius $R\approx 1100$ km. 

\section{Robustness against failures and attacks}

Here, we analyze and compare the robustness of the satellite and optical fiber models, considering the  behavior of the networks under random failures and targeted attacks on nodes and links \cite{albert2000error}. As a benchmark, we will also use the paradigmatic random \cite{erdos1959random} and scale-free networks \cite{Albert1999}. If there is a random failure in a node, all its connections are broken. Alternatively, one can think of a targeted attack breaking the nodes on the most connected sites of the network. Our goal is to determine how many nodes have to fail in network in order to break it apart.

To answer this we need to compute the ratio $N_G(f)/N_G(0)$ as a function of $f$ (number of removed nodes divided by the size $N$ of the network). The parameter $N_G(f)$ is the size of the giant cluster after we remove a fraction $f$ of nodes from the network, and $N_G(0)$ is the size of the largest cluster before any node removal. Analysing this ratio as a function of $f$ is the standard procedure in network science to obtain the robustness of the network under random failures (if we randomly remove the nodes) or target attack (if start removing the most connected nodes of the network)
 \cite{barabasi2016network,albert2000error,PhysRevLett.85.4626}.

\begin{figure*}[!t]
\begin{center}
\includegraphics[width=\textwidth]{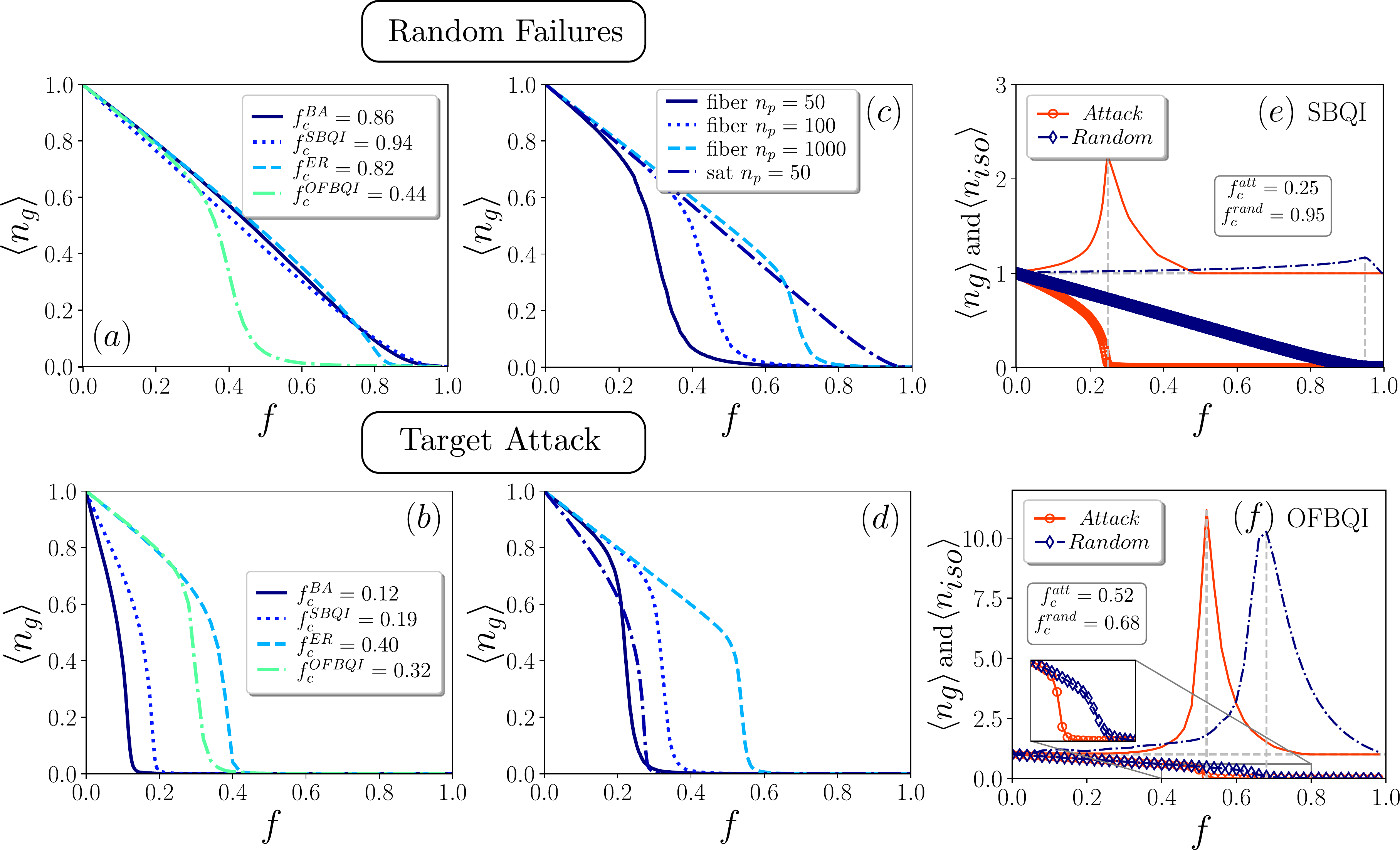}
\end{center}
\caption{\textbf{Quantum Network Robustness}. Comparison between the optical fiber and satellite models having also the Barabási-Albert (BA) (scale-free model) and the standard Erdos-Renyi random graph (ER) as benchmarks. We set $N=10000$ and, as standard in the literature, choose the parameters of the each model to get a network with $\langle k \rangle \simeq 6$. $(a)$ The SBQI is as robust as the BA model under random failures. A removal of $94\%$ of the nodes is necessary to break the network. On the other hand, the OFBQI is the least robust network. $(b)$ As expected, the ER model is the most robust under targeted attacks, with $f_c \sim 40\%$, followed by the OFBQI with $f_c=32\%$. The least robust is the BA $(f_c \sim 12\%)$ that is similar to SBQI model with $f_c \sim 19\%$. In figures $(c)$ and $(d)$ a comparison between the fiber and satellite models fixing $N=10000$ and $\rho=0.00022$ sites/km, varying $n_p = 50$, $100$ and $1000$ for OFBQI network and fixing $n_p=50$ for SBQI model. Even allowing for considerably more losses, the OFBQI is less robust than SBQI network under random failures and it becomes less robust as $n_p$ decreases. However, the SBQI model is very fragile against targeted attacks, while OFBQI becomes more robust for larger values of $n_p $ (for instance $n_p\geq100$). In  graphs $(e)-(f)$ we show the critical threshold $f_c$ for SBQI and OFBQI under random failures and target attacks fixing $\rho = 0.0002$ sites/km. $(e)$ SBQI, with $n_p = 50$, and $(f)$ OFBQI, with $n_p = 1000$, under random failures (blue) and target attacks (red). We compare $\langle n_g\rangle$ (blue diamonds/red circles) with  $\langle n_{iso}\rangle$ (blue dashed dotted line/red straight line) as a function of the fraction $f$ of removed nodes under random failures/targeted attacks. The peak of the $\langle n_{iso} \rangle $ coincides with $\langle n_g\rangle \sim 0$ indicating the value of the critical threshold $f_c$. 
}
\label{robustness_all}
\end{figure*}

In Fig. \ref{robustness_all} we show the average value of $N_G(f)/N_G(0)$, denoted by $\langle n_g\rangle$ for both random failures and targeted attacks. For random failures, the SBQI is significantly more robust than the OFBQI, as it can be seen comparing the critical values $f_c$ (for which $\langle n_g\rangle$ is approximately zero) and the faster decay rate of the OFBQI. Furthermore, we can see that the SBQI has a similar behaviour to a scale-free network, that is known to be robust against random failures \cite{albert2000error,failure}.

Due to the existence of hubs, it is well known that the scale-free model is less robust against targeted attacks as compared, for instance, with random networks. In the SBQI such hubs are also present and as shown in Fig. \ref{robustness_all} this leads to a smaller robustness in this situation.

\begin{figure*}[!t]
\begin{center}
\includegraphics[width=.7\textwidth]{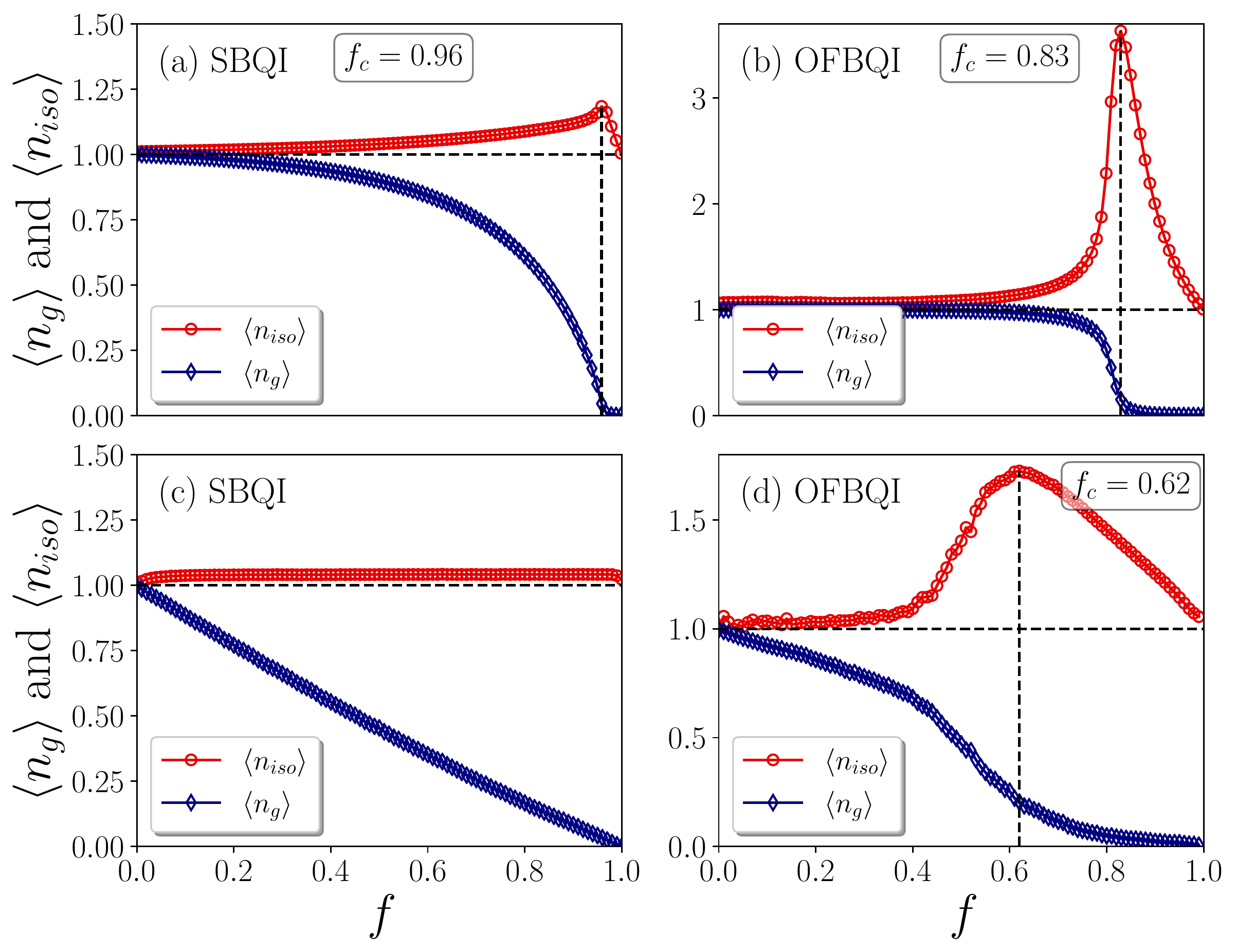}
\end{center}
\caption{\textbf{Robustness under link removal.} In these figures we show the the robustness of $(a)$ SBQI and $(b)$ under random link failures. The results were done for $\rho=0.0002$, $N=10000$ under $1000$ realizations. Below, we show the robustness of the models,  $(c)$ SBQI and $(d)$ OFBQI, by removing the minimum amount of links necessary to break the communication between two randomly chosen nodes. The results were done for $\rho = 0.0002$ sites/km, $N = 1000$ under $1000$ realizations. In both cases the $SBQI$ is more robust than $OFBQI$ network.}
\label{robustness_edge}
\end{figure*}

An alternative  to obtain the critical threshold $f_c$ is to compare $\langle n_g\rangle$ with the average size of the isolated clusters $\langle n_{iso} \rangle$ as a function of the fraction $f$ of removed nodes \cite{albert2000error}. 
As shown in Fig. ~\ref{robustness_all}(e) and (f) the peak of the $\langle n_{iso} \rangle $ coincides with $\langle n_g\rangle \sim 0$ indicating the value of the critical threshold $f_c$. From this point on, the network loses its communication capacity. 
As it can be seen in Fig. \ref{robustness_all} (f), under random failures, a failure of $68\%$ of nodes it is enough to bring down the communication capabilities of the OFBQI network, whereas for the SBQI  network (Fig.~\ref{robustness_all}(e)) almost all nodes of the network ($\sim 95\%$) would have to fail in order to break it down. In turn, under targeted attacks, the failure of only $\sim 25\%$ of the nodes, would be already enough to break the SBQI network (Fig.~\ref{robustness_all}(e)), while for OFBQI it is necessary that $\sim 52\%$ of the nodes fail (Fig.~\ref{robustness_all}(f)).

We also study two cases of link failure: 1) random link failures; 2) ''edge-cut'' attack. Notice that the first kind of failure would correspond to the sources of noise (decoherence etc). And the second kind would correspond to an attack choosing specifically the links necessary to break down all the paths between two nodes (breaking off the communication between them).
In the Figs.~\ref{robustness_edge}(a,b), we are fixing the parameters $\rho = 0.0002$ sites/km, $N = 10000$ under $1000$ realizations, and we show the robustness of the (a) $SBQI$ and (b) $OFBQI$ models under random link failures. By comparing the random link failure with the case of the random node failure, we can see that for the SBQI network the critical thresholds $f_c$ are equivalent.  This result is known for networks where the probability to link two sites does not strongly depend on distances between them (for instance, Erdos-Renyi \cite{erdos1959random} and Barabási-Albert \cite{Albert1999} networks). As discussed in the text, in the SBQI model the distance between the sites is almost irrelevant in the link probability, explaining this result. On the contrary, the OFBQI model displays higher robustness under random link failures as compared to node failures. In conclusion, also for random link failures the SBQI is still much more robust than the OFBQI.
 
In the second case showed in Figs.~\ref{robustness_edge}(c,d) (''edge-cut'' attack), we compare the networks by removing the minimum amount of links necessary to fully disconnect two randomly chosen nodes. As can be seen in the figures (c) SBQI and (d) OFBQI, under this kind of attack, the SBQI model is again more robust than OFBQI model. We can see in figure (c) that there is no critical threshold from what $N_G/N = 0$ and the network becomes disconnected. However, the OFBQI network becomes more fragile under this kind of attack.

In summary, under link failures, the SBQI is again more robust than OFBQI. We must highlight, however, that the results regarding link failures are preliminary. The bottleneck of the analysis is the computational cost to consider very large networks. For this reason we  have considered a relatively small $N$ as compared to the node failure case.

\section{Discussion}

\normalsize We have studied the network properties of a quantum internet assuming that the links interconnecting the different nodes are quantum channels mediated by a satellite with down-link communication with stations on the ground. We have shown that such networks display hubs, \ie nodes that have a big number of connections, naturally present closer to the satellite. These nodes have the effect of making the typical networks distances and diameter small. This leads to a clear advantage of such networks in entanglement distribution, as compared to networks based on optical fibers \cite{samurai1}. The presence of hubs also make the network more robust against random node and link failures, since these nodes also have the capability of holding the network together after a considerable number of nodes are removed. However, if the attacks are targeted to destroy the hubs, the network are dismantled pretty easily. This highlights the need for a special protection of these nodes. 

Our results provide a useful guide for the development of future quantum networks. Together with the network based on optical fibers \cite{samurai1}, our results can be seen as the first step towards more complicated and realistic models.
It would be interesting to consider non-uniform distribution of nodes that simulates, for instance, the fact that
big cities typically concentrate more nodes than rural areas. We expect
that this situation could lead to the appearance of communities. Other important line of research could be to study quantum features of the transmitted photons, such as coherence and entanglement, and how these features impact the usefulness of the network for specific protocols. Finally, we believe that future quantum networks will be hybrid, using simultaneously optical fibers and satellites in the most efficient way.
%where one could, for instance, analise hybrid networks employing both fiber and satellite communication, consider quantum features of the communicating photons such as coherence and entanglement, study the performance of specific protocols \cite{pirandola2019end,rigovacca2018versatile,bauml2020linear,azuma2016fundamental,takeoka2019multipartite}, and consider different topologies among many other open questions. 

\acknowledgements
We acknowledge the John Templeton Foundation via the Grant Q-CAUSAL No. 61084, the Serrapilheira Institute (Grant No. Serra-1708-15763), the Brazilian National Council for Scientific and Technological Development (CNPq) via the National Institute for Science and Technology on Quantum Information (INCT-IQ) and Grants No. $423713/2016-7$, No. 307172/2017-1 and No. 406574/2018-9, the Brazilian agencies MCTIC and MEC. DC acknowledges the Ramon y Cajal fellowship, the Spanish MINECO (QIBEQI FIS2016-80773-P, Severo Ochoa SEV-2015-0522), Fundacio Cellex, and the Generalitat de Catalunya (SGR 1381 and CERCA Programme). AC acknowledges UFAL for a paid license for scientific cooperation at UFRN. We thank the High Performance Computing Center (NPAD/UFRN) for providing computational resources.

\bibliography{Ref}

\end{document}